\documentclass{ws-ijmpd}
\usepackage[super,compress]{cite}
\usepackage{lmodern}
\usepackage{color}
\usepackage{amsmath}
\usepackage{amssymb}
\usepackage{graphicx}
\usepackage{floatrow}
\usepackage[unicode=true,pdfusetitle,
 bookmarks=true,bookmarksnumbered=false,bookmarksopen=false,
 breaklinks=false,pdfborder={0 0 1},colorlinks=true]
 {hyperref}

\begin{document}

\markboth{Nilok Bose, A. S. Majumdar}
{Study of cosmic backreaction on the future evolution of an accelerating
universe using multiple domains}

\title{Study of cosmic backreaction on the future evolution of an accelerating
universe using multiple domains}

\author{Nilok Bose$^*$ and A. S. Majumdar }

\address{S. N. Bose National Centre for Basic Sciences,\\
Block JD, Sector III, Salt Lake,\\
Kolkata 700098, India\\
$^*$E-mail: nilok@bose.res.in}

\maketitle

\begin{abstract}
We investigate the effect of backreaction due to inhomogeneities on
the evolution of the present Universe by considering the Universe
to be partitioned into multiple domains within the Buchert framework.
Taking the observed present acceleration of the universe as an essential
input, we study the effect of inhomogeneities on the future evolution.
We find that the backreaction from inhomogeneities causes the acceleration
to slow down in the future for a range of initial configurations and
model parameters, and even lead in certain cases to the emergence
of a future decelerating epoch. We consider two different partitioning
of the Universe and perform a comparative analysis for the two cases
on the behaviour of the acceleration and backreaction of the Universe.
\end{abstract}

\ccode{PACS numbers: 98.80.-k, 95.36.+x, 98.80.Es}

\section{Introduction}

By now the present acceleration of the Universe is quite well established
observationally \cite{perlmutter}, and the cause for it is attributed
to a mysterious component called Dark Energy whose true nature is
still unknown to us. A cosmological constant serves as the simplest
possible explanation for the current acceleration but it is endowed
with several conceptual problems \cite{weinberg}. There is no dearth
of innovative ideas to explain the current acceleration \cite{sahni}
but our present state of knowledge offers no clear indication on the
nature of the big rip that our Universe seems to be headed for.

We know from observations that the Universe is inhomogeneous up to
the scales of super-clusters of galaxies, which suggests that some
modification is needed for our standard cosmological framework that
is based on the globally smooth Friedmann\textendash{}Robertson\textendash{}Walker
(FRW) metric. Since the equations of general relativity are non-linear
in nature therefore if we take the global average of the Einstein
tensor then those results are very unlikely to be the same as that
obtained by taking the average over all the different local metrics
and then computing the global Einstein tensor. This realization has
lead to the investigation of the question of how backreaction originating
from density inhomogeneities could modify the evolution of the universe
as described by the background FRW metric at large scales.

Several approaches have been developed to facilitate the study of
the effects of inhomogeneous matter distribution on the evolution
of the Universe, such as Zalaletdinov's fully covariant macroscopic
gravity \cite{zala}; Buchert's approach of averaging the scalar parts
of Einstein's equations \cite{buchert1,buchert2}, the perturbation
techniques proposed by Kolb et. al. \cite{kolb1}, and recently another
procedure based on light-cone averaging has been proposed \cite{gasperini2}.
It has also been argued, by using the framework developed by Buchert,
that backreaction from inhomogeneities could be responsible for the
current acceleration of the Universe \cite{rasanen1}. A different
perspective of the Buchert framework developed by Wiltshire \cite{wiltshire1}
also leads to an apparent acceleration due to the different lapse
of time in underdense and overdense regions.

Since inhomogeneities lead to the formation of structures in the present
era therefore the effects of backreaction will also gain strength
in the present era, and hence if backreaction is indeed responsible
for the current accelerated expansion then it will provide an interesting
platform to study this issue, without invoking additional physics.
While there is some debate on the impact of inhomogeneities on observables
of an overall homogeneous FRW model \cite{ishibashi,paranajpe,kolb2},
and there are also similar questions with regard to the magnitude
of backreaction modulated by the effect of shear between overdense
and underdense regions \cite{mattson}, recent studies \cite{buchert1,buchert2,buchert3,rasanen1,kolb1,wiltshire1,kolb2,weigand}
have provided a strong motivation for exploring further the role of
inhomogeneities on the evolution of the present Universe. Nonetheless
further investigations with the input of data from future observational
probes are required for a conclusive picture to emerge.

Recently we studied the backreaction scenario within the Buchert framework
using a simple two-scale model \cite{bose}. Our model indicated the
possibility of a transition to a future decelerated era. The Buchert
framework has been further extended in~\refcite{weigand} so as to facilitate
the study of backreaction in the case where the universe is considered
to be divided into multiple domains and subdomains. But the model
that was considered in~\refcite{weigand} was a simple model consisting
of one overdense subdomain and one underdense subdomain, like that
used in~\refcite{bose}. Such a simple model is attractive because it
simplifies the evolution equations and eases the process of understanding
the effect of backreaction on the evolution of the Universe. But the
real Universe cannot be partitioned simply into one overdense subdomain
and one underdense subdomain. To the best of our knowledge, so far
there has been no study on the effect of backreaction from inhomogeneities
by considering multiple subdomains. Using the formalism proposed in~\refcite{weigand}, 
in the present paper we improve upon our previous
two-scale model and consider the Universe as a global domain $\mathcal{D}$
which is partitioned into multiple overdense and underdense regions,
and all these subdomains are taken to evolve differently to each other.
This is done in order to recreate the real Universe as much as possible
in our simple model. Our aim here is to study the future evolution
of the Universe by taking into consideration its current accelerated
expansion. The accelerated expansion of the Universe can be assumed
to be caused by backreaction or any other mechanism \cite{sahni}.
We consider two different partitioning cases of the Universe and explore
the future evolution of the Universe for these two cases and then
perform a comparative analysis for the two.

The paper is organized as follows. In Section II we briefly recapitulate
the essential details of the Buchert framework \cite{buchert1,buchert2,buchert3}
including the evolution equations when the Universe is partitioned
into multiple subdomains. Next, in Section III we follow the approach
of~\refcite{weigand} and investigate the future evolution of the Universe
assuming its present stage of global acceleration. Subsequently, in
Section IV we perform a quantitative comparison of various dynamical
features of our model for the two partitioning cases that we consider.
Finally, we summarize our results and make some concluding remarks
in Section V.

\section{The Backreaction Framework}

\subsection{Averaged Einstein equations}

In the framework developed by Buchert \cite{buchert1,buchert2,buchert3,weigand}
for the Universe filled with an irrotational fluid of dust, the space\textendash{}time
is foliated into flow-orthogonal hypersurfaces featuring the line-element
\begin{equation}
ds^{2}=-dt^{2}+g_{ij}dX^{i}dX^{j}
\end{equation}
where the proper time $t$ labels the hypersurfaces and $X^{i}$ are
Gaussian normal coordinates (locating free-falling fluid elements
or generalized fundamental observers) in the hypersurfaces, and $g^{ij}$
is the full inhomogeneous three metric of the hypersurfaces of constant
proper time. The framework is applicable to perfect fluid matter models.

For a compact spatial domain $\mathcal{D}$, comoving with the fluid,
there is one fundamental quantity characterizing it and that is its
volume. This volume is given by 
\begin{equation}
|\mathcal{D}|_{g}=\int_{\mathcal{D}}d\mu_{g}
\end{equation}
where $d\mu_{g}=\sqrt{^{(3)}g(t,X^{1},X^{2},X^{3})}dX^{1}dX^{2}dX^{3}$.
From the domain's volume one may define a scale-factor 
\begin{equation}
a_{\mathcal{D}}(t)=\left(\frac{|\mathcal{D}|_{g}}{|\mathcal{D}_{i}|_{g}}\right)^{1/3}
\end{equation}
that encodes the average stretch of all directions of the domain.

Using the Einstein equations, with a pressure-less fluid source, we
get the following equations \cite{buchert1,buchert3,weigand} 
\begin{eqnarray}
3\frac{\ddot{a}_{\mathcal{D}}}{a_{\mathcal{D}}} & = & -4\pi G\left\langle \rho\right\rangle _{\mathcal{D}}+\mathcal{Q}_{\mathcal{D}}+\Lambda\label{eq:1a}\\
3H_{\mathcal{D}}^{2} & = & 8\pi G\left\langle \rho\right\rangle _{\mathcal{D}}-\frac{1}{2}\mathcal{\left\langle R\right\rangle }_{\mathcal{D}}-\frac{1}{2}\mathcal{Q}_{\mathcal{D}}+\Lambda\label{eq:1b}\\
0 & = & \partial_{t}\left\langle \rho\right\rangle _{\mathcal{D}}+3H_{\mathcal{D}}\left\langle \rho\right\rangle _{\mathcal{D}}\label{eq:1c}
\end{eqnarray}
Here the average of the scalar quantities on the domain $\mathcal{D}$
is defined as, 
\begin{equation}
\left\langle f\right\rangle {}_{\mathcal{D}}(t)=\frac{\int_{\mathcal{D}}f(t,X^{1},X^{2},X^{3})d\mu_{g}}{\int_{\mathcal{D}}d\mu_{g}}=|\mathcal{D}|_{g}^{-1}\int_{\mathcal{D}}fd\mu_{g}\label{eq:2}
\end{equation}
and where $\rho$, $\mathcal{R}$ and $H_{\mathcal{D}}$ denote the
local matter density, the Ricci-scalar of the three-metric $g_{ij}$,
and the domain dependent Hubble rate $H_{\mathcal{D}}=\dot{a}_{\mathcal{D}}/a_{\mathcal{D}}$
respectively. The kinematical backreaction $\mathcal{Q_{D}}$ is defined
as 
\begin{equation}
\mathcal{Q_{D}}=\frac{2}{3}\left(\left\langle \theta^{2}\right\rangle _{\mathcal{D}}-\left\langle \theta\right\rangle _{\mathcal{D}}^{2}\right)-2\sigma_{\mathcal{D}}^{2}\label{eq:3}
\end{equation}
where $\theta$ is the local expansion rate and $\sigma^{2}=1/2\sigma_{ij}\sigma^{ij}$
is the squared rate of shear. It should be noted that $H_{\mathcal{D}}$
is defined as $H_{\mathcal{D}}=1/3\left\langle \theta\right\rangle _{\mathcal{D}}$.
$\mathcal{Q_{D}}$ encodes the departure from homogeneity and for
a homogeneous domain its value is zero.

One also has an integrability condition that is necessary to yield
\eqref{eq:1b} from \eqref{eq:1a} and that relation reads as 
\begin{equation}
\frac{1}{a_{\mathcal{D}}^{6}}\partial_{t}\left(a_{\mathcal{D}}^{6}\mathcal{Q}_{\mathcal{D}}\right)+\frac{1}{a_{\mathcal{D}}^{2}}\partial_{t}\left(a_{\mathcal{D}}^{2}\mathcal{\left\langle R\right\rangle }_{\mathcal{D}}\right)=0
\end{equation}
From this equation we see that the evolution of the backreaction term,
and hence extrinsic curvature inhomogeneities, is coupled to the average
intrinsic curvature. Unlike the FRW evolution equations where the
curvature term is restricted to an $a_{\mathcal{D}}^{-2}$ behaviour,
here it is more dynamical because it can be any function of $a_{\mathcal{D}}$.

\subsection{Separation formulae for arbitrary partitions}

The ``global'' domain $\mathcal{D}$ is assumed to be separated
into subregions $\mathcal{F}_{\ell}$ , which themselves consist of
elementary space entities $\mathcal{F}_{\ell}^{(\alpha)}$ that may
be associated with some averaging length scale. In mathematical terms
$\mathcal{D}=\cup_{\ell}\mathcal{F}_{\ell}$, where $\mathcal{F}_{\ell}=\cup_{\alpha}\mathcal{F}_{\ell}^{(\alpha)}$
and $\mathcal{F}_{\ell}^{(\alpha)}\cap\mathcal{F}_{m}^{(\beta)}=\emptyset$
for all $\alpha\neq\beta$ and $\ell\neq m$. The average of the scalar
valued function $f$ on the domain $\mathcal{D}$ \eqref{eq:2} may
then be split into the averages of $f$ on the subregions $\mathcal{F}_{\ell}$
in the form, 
\begin{equation}
\left\langle f\right\rangle _{\mathcal{D}}=\underset{\ell}{\sum}|\mathcal{D}|_{g}^{-1}\underset{\alpha}{\sum}\int_{\mathcal{F}_{\ell}^{(\alpha)}}fd\mu_{g}=\underset{\ell}{\sum}\lambda_{\ell}\left\langle f\right\rangle _{\mathcal{F}_{\ell}}
\end{equation}
where $\lambda_{\ell}=|\mathcal{F}_{\ell}|_{g}/|\mathcal{D}|_{g}$,
is the volume fraction of the subregion $\mathcal{F}_{\ell}$. The
above equation directly provides the expression for the separation
of the scalar quantities $\rho$, $\mathcal{R}$ and $H_{\mathcal{D}}=1/3\left\langle \theta\right\rangle _{\mathcal{D}}$.
However, $\mathcal{Q_{D}}$, as defined in \eqref{eq:3}, does not
split in such a simple manner due to the $\left\langle \theta\right\rangle _{\mathcal{D}}^{2}$
term. Instead the correct formula turns out to be 
\begin{equation}
\mathcal{Q_{D}}=\underset{\mathcal{D}}{\mathcal{\sum}}\lambda_{\ell}\mathcal{Q}_{\ell}+3\underset{\ell\neq m}{\sum}\lambda_{\ell}\lambda_{m}\left(H_{\ell}-H_{m}\right)^{2}\label{eq:4}
\end{equation}
where $\mathcal{Q}_{\ell}$ and $H_{\ell}$ are defined in $\mathcal{F}_{\ell}$
in the same way as $\mathcal{Q}_{\mathcal{D}}$ and $H_{\mathcal{D}}$
are defined in $\mathcal{D}$. The shear part $\left\langle \sigma^{2}\right\rangle _{\mathcal{F}_{\ell}}$
is completely absorbed in $\mathcal{Q}_{\ell}$ , whereas the variance
of the local expansion rates $\left\langle \theta^{2}\right\rangle _{\mathcal{D}}-\left\langle \theta\right\rangle _{\mathcal{D}}^{2}$
is partly contained in $\mathcal{Q}_{\ell}$ but also generates the
extra term $3\sum_{\ell\neq m}\lambda_{\ell}\lambda_{m}\left(H_{\ell}-H_{m}\right)^{2}$.
This is because the part of the variance that is present in $\mathcal{Q}_{\ell}$,
namely $\left\langle \theta^{2}\right\rangle _{\mathcal{\mathcal{F}_{\ell}}}-\left\langle \theta\right\rangle _{\mathcal{F}_{\ell}}^{2}$
only takes into account points inside $\mathcal{F}_{\ell}$. To restore
the variance that comes from combining points of $\mathcal{F}_{\ell}$
with others in $\mathcal{F}_{m}$, the extra term containing the averaged
Hubble rate emerges. Note here that the above formulation of the backreaction
holds in the case when there is no interaction between the overdense
and the underdense subregions.

Analogous to the scale-factor for the global domain, a scale-factor
$a_{\ell}$ for each of the subregions $\mathcal{F}_{\ell}$ can be
defined such that $|\mathcal{D}|_{g}=\sum_{\ell}|\mathcal{F}_{\ell}|_{g}$,
and hence, 
\begin{equation}
a_{\mathcal{D}}^{3}=\sum_{\ell}\lambda_{\ell_{i}}a_{\ell}^{3}\label{globscale}
\end{equation}
where $\lambda_{\ell_{i}}=|\mathcal{F}_{\ell_{i}}|_{g}/|\mathcal{D}_{i}|_{g}$
is the initial volume fraction of the subregion $\mathcal{F}_{\ell}$.
If we now twice differentiate this equation with respect to the foliation
time and use the result for $\dot{a}_{\ell}$ from \eqref{eq:1b},
we then get the expression that relates the acceleration of the global
domain to that of the sub-domains: 
\begin{equation}
\frac{\ddot{a}_{\mathcal{D}}}{a_{\mathcal{D}}}=\underset{\ell}{\sum}\lambda_{\ell}\frac{\ddot{a}_{\ell}(t)}{a_{\ell}(t)}+\underset{\ell\neq m}{\sum}\lambda_{\ell}\lambda_{m}\left(H_{\ell}-H_{m}\right)^{2}\label{eq:5}
\end{equation}

\section{Future evolution within the Buchert framework}

We will now explore the future evolution of the Universe after the
current stage of acceleration sets in. It is not necessary to assume
that the acceleration is caused by backreaction \cite{rasanen1,weigand}.
For the purpose of our present analysis, it suffices to consider the
observed accelerated phase of the universe \cite{seikel} that could
occur due to any of a variety of mechanisms \cite{sahni}.

Following the Buchert framework \cite{buchert1,weigand} as discussed
above, the global domain $\mathcal{D}$ is divided into multiple domains.
We consider $\mathcal{D}$ to be partitioned into equal numbers of
overdense and underdense domains. We label all overdense domains as
$\mathcal{M}$ (called `Wall') and all underdense domains as $\mathcal{E}$
(called `Void'), such that $\mathcal{D}=\left(\mathcal{\cup_{\mathit{j}}}\mathcal{M}^{j}\right)\cup\left(\cup_{\mathit{j}}\mathcal{E}^{j}\right)$.
In this case one obtains $H_{\mathcal{D}}=\sum_{j}\lambda_{\mathcal{M}_{j}}H_{\mathcal{M}_{j}}+\sum_{j}\lambda_{\mathcal{E}_{j}}H_{\mathcal{E}_{j}}$,
with similar expressions for $\left\langle \rho\right\rangle _{\mathcal{D}}$
and $\left\langle \mathcal{R}\right\rangle _{\mathcal{D}}$ and also
$\sum_{j}\lambda_{j}=1$. For such a partitioning the global acceleration
\eqref{eq:5} can be written as 

\begin{eqnarray}
\frac{\ddot{a}_{\mathcal{D}}}{a_{\mathcal{D}}} & = & \sum_{j}\lambda_{\mathcal{M}_{j}}\frac{\ddot{a}_{\mathcal{M}_{j}}}{a_{\mathcal{M}_{j}}}+\sum_{j}\lambda_{\mathcal{E}_{j}}\frac{\ddot{a}_{\mathcal{E}_{j}}}{a_{\mathcal{E}_{j}}}\nonumber \\
 &  & +\underset{j\neq k}{\sum}\lambda_{\mathcal{M}_{j}}\lambda_{\mathcal{M}_{k}}(H_{\mathcal{M}_{j}}-H_{\mathcal{M}_{k}})^{2}\nonumber \\
 &  & +\underset{j\neq k}{\sum}\lambda_{\mathcal{E}_{j}}\lambda_{\mathcal{E}_{k}}(H_{\mathcal{E}_{j}}-H_{\mathcal{E}_{k}})^{2}\nonumber \\
 &  & +2\underset{j,\, k}{\sum}\lambda_{\mathcal{M}_{j}}\lambda_{\mathcal{E}_{k}}(H_{\mathcal{M}_{j}}-H_{\mathcal{E}_{k}})^{2}\label{eq:6}
\end{eqnarray}
We assume that the scale-factors of the regions $\mathcal{E}^{j}$
and $\mathcal{M}^{j}$ are, respectively, given by $a_{\mathcal{E}_{j}}=c_{\mathcal{E}_{j}}t^{\alpha_{j}}$
and $a_{\mathcal{M}_{j}}=c_{\mathcal{M}_{j}}t^{\beta_{j}}$ where
$\alpha_{j}$, $\beta_{j}$, $c_{\mathcal{E}_{j}}$ and $c_{\mathcal{M}_{j}}$
are constants. The volume fraction of the subdomain $\mathcal{M}^{j}$
is given by $\lambda_{\mathcal{M}_{j}}=\frac{|\mathcal{M}^{j}|_{g}}{|\mathcal{D}|_{g}}$,
which can be rewritten in terms of the corresponding scale factors
as $\lambda_{\mathcal{M}_{j}}=\frac{a_{\mathcal{M}_{j}}^{3}|\mathcal{M}_{i}^{j}|_{g}}{a_{\mathcal{D}}^{3}|\mathcal{D}_{i}|_{g}}$,
and similarly for the $\mathcal{E}^{j}$ subdomains. We therefore
find that the global acceleration equation \eqref{eq:6} becomes 
\begin{eqnarray}
\frac{\ddot{a}_{\mathcal{D}}}{a_{\mathcal{D}}} & = & \sum_{j}\frac{g_{\mathcal{M}_{j}}^{3}t^{3\beta_{j}}}{a_{\mathcal{D}}^{3}}\frac{\beta_{j}(\beta_{j}-1)}{t^{2}}+\sum_{j}\frac{g_{\mathcal{E}_{j}}^{3}t^{3\alpha_{j}}}{a_{\mathcal{D}}^{3}}\frac{\alpha_{j}(\alpha_{j}-1)}{t^{2}}\nonumber \\
 &  & +\underset{j\neq k}{\sum}\frac{g_{\mathcal{M}_{j}}^{3}t^{3\beta_{j}}}{a_{\mathcal{D}}^{3}}\frac{g_{\mathcal{M}_{k}}^{3}t^{3\beta_{k}}}{a_{\mathcal{D}}^{3}}\left(\frac{\beta_{j}}{t}-\frac{\beta_{k}}{t}\right)^{2}\nonumber \\
 &  & +\underset{j\neq k}{\sum}\frac{g_{\mathcal{E}_{j}}^{3}t^{3\alpha_{j}}}{a_{\mathcal{D}}^{3}}\frac{g_{\mathcal{E}_{k}}^{3}t^{3\alpha_{k}}}{a_{\mathcal{D}}^{3}}\left(\frac{\alpha_{j}}{t}-\frac{\alpha_{k}}{t}\right)^{2}\nonumber \\
 &  & +2\underset{j,\, k}{\sum}\frac{g_{\mathcal{M}_{j}}^{3}t^{3\beta_{j}}}{a_{\mathcal{D}}^{3}}\frac{g_{\mathcal{E}_{k}}^{3}t^{3\alpha_{k}}}{a_{\mathcal{D}}^{3}}\left(\frac{\beta_{j}}{t}-\frac{\alpha_{k}}{t}\right)^{2}\label{eq:19}
\end{eqnarray}
where $g_{\mathcal{M}_{j}}^{3}=\frac{\lambda_{\mathcal{M}_{j_{0}}}a_{\mathcal{D}_{0}}^{3}}{t_{0}^{3\beta_{j}}}$
and $g_{\mathcal{E}_{j}}^{3}=\frac{\lambda_{\mathcal{E}_{j_{0}}}a_{\mathcal{D}_{0}}^{3}}{t_{0}^{3\alpha_{j}}}$
are constants. 

\begin{figure}
\includegraphics[width=8.3cm]{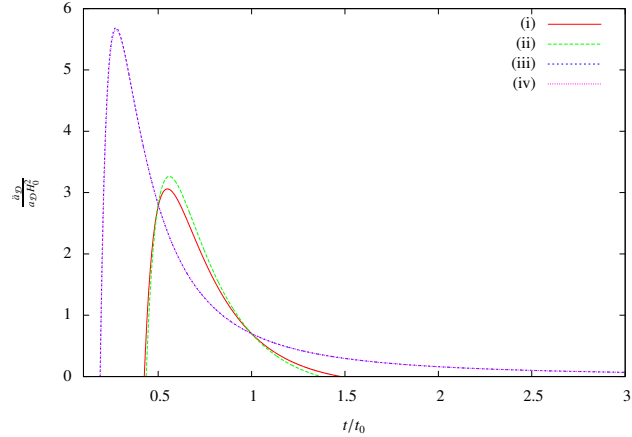}

\caption{The dimensionless global acceleration parameter $\frac{\ddot{a}_{\mathcal{D}}}{a_{\mathcal{D}}H_{0}^{2}}$,
plotted vs. time (in units of $t/t_{0}$ with $t_{0}$ being the current
age of the Universe ). In curves (i) and (ii) the value of $\alpha$
is in the range 0.990 - 0.999 and that of $\beta$ is in the range
0.58 - 0.60. In curves (iii) and (iv) the value of $\alpha$ is in
the range 1.02 - 1.04 and that of $\beta$ is in the range 0.58 -
0.60 }

\end{figure}

\begin{figure}
\includegraphics[width=8.3cm]{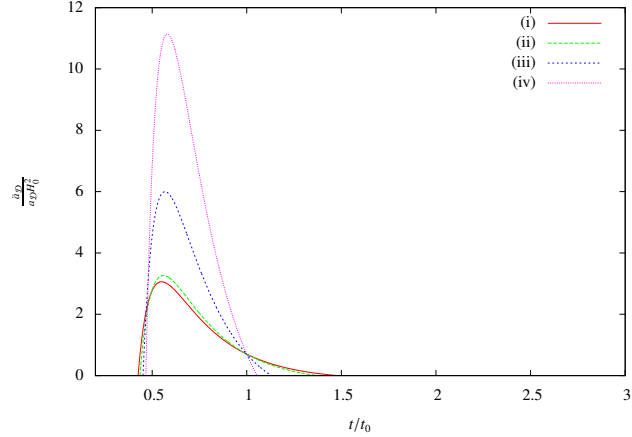}

\caption{Here also $\frac{\ddot{a}_{\mathcal{D}}}{a_{\mathcal{D}}H_{0}^{2}}$
is plotted vs. time. In curves (i) and (ii) the value of $\alpha$
lies in the range 0.990 - 0.999 and that of $\beta$ is in the range
0.58 - 0.60. For curves (iii) and (iv) the value of $\alpha$ is in
the range 0.990 - 0.999 and that of $\beta$ is in the range 0.55
- 0.65}

\end{figure}

We will perform a comparative study of two cases where (i) the global
domain $\mathcal{D}$ is considered to be divided into 50 overdense
and underdense subdomains each and (ii) 100 overdense and underdense
subdomains each. We will obtain numerical solutions of equation \eqref{eq:19}
for various ranges of parameter values. In order to do this we will
consider the range of values for the parameters $\alpha_{j}$ and
$\beta_{j}$ as a Gaussian distribution, which is of the form $\frac{1}{\sigma\sqrt{2\pi}}\exp\left[-\frac{\left(x-\mu\right)^{2}}{2\sigma^{2}}\right]$,
where $\sigma$ is the standard deviation and $\mu$ is the mean (the
range of values corresponds to the full width at half maximum of the
distribution). We will also assign values for the volume fractions
$\lambda_{\mathcal{M}_{j}}$ and $\lambda_{\mathcal{E}_{j}}$ based
on a Gaussian distribution and impose the restriction that the total
volume fraction of all the overdense subdomains at present time should
be $0.09$, a value that has been determined through numerical simulations
in the earlier literature \cite{weigand}. Note here that using our
ansatz for the subdomain scale factors one may try to determine the
global scale factor through Eq.\eqref{globscale}. In order to do
so, one needs to know the inital volume fractions $\lambda_{\ell_{i}}$
which are in turn related to the $c_{\mathcal{E}_{j}}$ and $c_{\mathcal{M}_{j}}$.
However, in our approach based upon the Buchert framework \cite{buchert1,buchert2,weigand}
we do not need to determine $c_{\mathcal{E}_{j}}$ and $c_{\mathcal{M}_{j}}$,
but instead, obtain from Eq.\eqref{eq:19} the global scale factor
numerically by the method of recursive iteration, using as an `initial
condition' the observational constraint $q_{0}=-0.7$, where $q_{0}$
is the current value of the deceleration parameter. The expression
for $q_{0}$ is a completely analytic function of $\alpha_{j}$, $\beta_{j}$
and $t_{0}$, but since we are studying the effect of inhomogeneities
therefore the Universe cannot strictly be described based on a FRW
model and hence the current age of the Universe ($t_{0}$) cannot
be fixed based on current observations which use the FRW model to
fix the age. Instead for each combination of values of the parameters
$\alpha_{j}$ and $\beta_{j}$ we find out the value of $t_{0}$ for
our model from \eqref{eq:19} by taking $q_{0}=-0.7$.

It is of interest to study separately the behaviour of the backreaction
term in the Buchert model \cite{buchert1,weigand}. The backreaction
$\mathcal{Q}_{\mathcal{D}}$ is obtained from \eqref{eq:1a} to be
\begin{equation}
\mathcal{Q}_{\mathcal{D}}=3\frac{\ddot{a}_{\mathcal{D}}}{a_{\mathcal{D}}}+4\pi G\left\langle \rho\right\rangle _{\mathcal{D}}\label{eq:14}
\end{equation}
Note that we are not considering the presence of any cosmological
constant $\Lambda$ as shown in \eqref{eq:1a}. We can assume that
$\left\langle \rho\right\rangle _{\mathcal{D}}$ behaves like the
matter energy density, i.e. $\left\langle \rho\right\rangle _{\mathcal{D}}=\frac{c_{\rho}}{a_{\mathcal{D}}^{3}}$,
where $c_{\rho}$ is a constant. Now, observations tell us that the
current matter energy density fraction (baryonic and dark matter)
is about 27\% and that of dark energy is about 73\%. Assuming the
dark energy density to be of the order of $10^{-48}\left(GeV\right)^{4}$,
we get $\rho_{\mathcal{D}_{0}}\backsimeq3.699\times10^{-49}\left(GeV\right)^{4}$.
Thus, using the values for the global acceleration computed numerically,
the future evolution of the backreaction term $\mathcal{Q}_{\mathcal{D}}$
can also be computed (see Figs. 3 and 4, where we have plotted the
backreaction density fraction $\Omega_{\mathcal{Q}}^{\mathcal{D}}=-\frac{\mathcal{Q}_{\mathcal{D}}}{6H_{\mathcal{D}}^{2}}$).

\begin{figure}
\includegraphics[width=8.3cm]{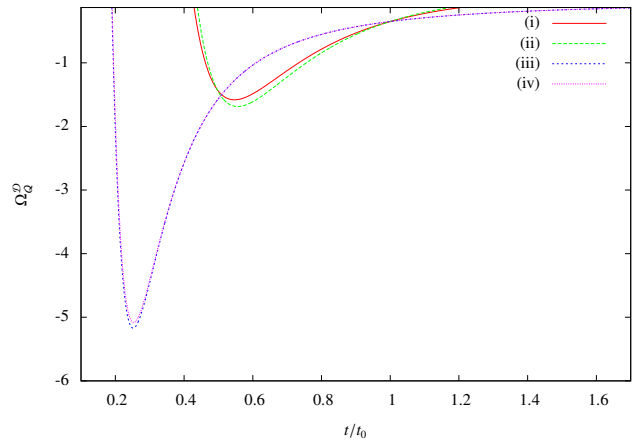}

\caption{Global backreaction density $\Omega_{\mathcal{Q}}^{\mathcal{D}}$
plotted vs. time (in units of $t/t_{0}$). In curves (i) and (ii)
the value of$\alpha$ is in the range 0.990 - 0.999 and that of $\beta$
is in the range 0.58 - 0.60. In curves (iii) and (iv) the value of
$\alpha$ is in the range 1.02 - 1.04 and that of $\beta$ is in the
range 0.58 - 0.60}

\end{figure}

\begin{figure}
\includegraphics[width=8.3cm]{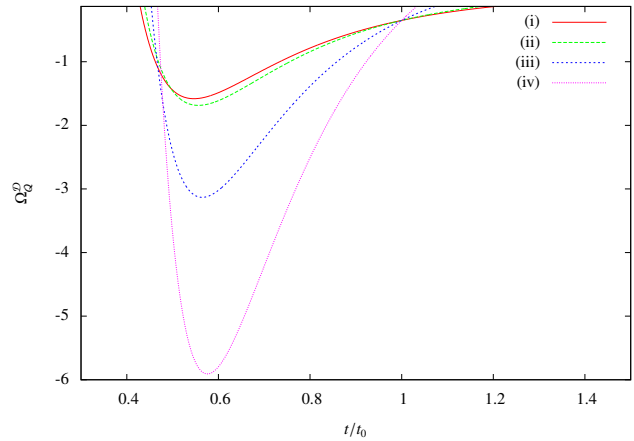}

\caption{Here also $\Omega_{\mathcal{Q}}^{\mathcal{D}}$ is plotted vs. time.
In curves (i) and (ii) the value of $\alpha$ lies in the range 0.990
- 0.999 and that of $\beta$ is in the range 0.58 - 0.60. For curves
(iii) and (iv) the value of $\alpha$ is in the range 0.990 - 0.999
and that of $\beta$ is in the range 0.55 - 0.65}

\end{figure}

\section{Discussions}

Let us now compare the nature of acceleration of the Universe for
the two cases described in the previous section. The global acceleration
for the two cases have been plotted in Figs. 1 and 2. In both the
figures curves (i) and (iii) are for the case where $\mathcal{D}$
is partitioned into 50 overdense and underdense subdomains each, and
curves (ii) and (iv) correspond to the case where $\mathcal{D}$ is
partitioned into 100 overdense and underdense subdomains each. The
values for the expansion parameters $\beta_{j}$ of the overdense
subdomains is taken to lie between $1/2$ and $2/3$ since the expansion
is assumed to be faster than in the radiation dominated case, and
is upper limited by the value for matter dominated expansion. In Fig.
1 the behaviour of global acceleration is shown for values of $\alpha_{j}<1$
and also $\alpha_{j}>1$, keeping the range of values of $\beta_{j}$
quite narrow and also the same for all four curves. We have kept the
value of $\alpha_{j}$ close to $1$ when $\alpha_{j}<1$ because
if $\alpha_{j}$ is less than a certain value, which depends on the
value of $\beta_{j}$, then the acceleration becomes undefined as
we do not get real solutions from \eqref{eq:19}. In order to demonstrate
this fact analytically let us consider a toy model where $\mathcal{D}$
is divided into one overdense subdomain $\mathcal{M}$ and one underdense
subdomain $\mathcal{E}$. In this case \eqref{eq:19} can be written
as

\begin{eqnarray}
\frac{\ddot{a}_{\mathcal{D}}}{a_{\mathcal{D}}} & = & \frac{g_{\mathcal{M}}^{3}t^{3\beta}}{a_{\mathcal{D}}^{3}}\frac{\beta(\beta-1)}{t^{2}}+\frac{g_{\mathcal{E}}^{3}t^{3\alpha}}{a_{\mathcal{D}}^{3}}\frac{\alpha(\alpha-1)}{t^{2}}\nonumber \\
 &  & +2\frac{g_{\mathcal{M}}^{3}t^{3\beta}}{a_{\mathcal{D}}^{3}}\frac{g_{\mathcal{E}}^{3}t^{3\alpha}}{a_{\mathcal{D}}^{3}}\left(\frac{\beta}{t}-\frac{\alpha}{t}\right)^{2}
\end{eqnarray}

Now we must have $\lambda_{\mathcal{M}}+\lambda_{\mathcal{E}}=1$,
so we can write $\frac{g_{\mathcal{E}}^{3}t^{3\beta}}{a_{\mathcal{D}}^{3}}=1-\frac{g_{\mathcal{M}}^{3}t^{3\beta}}{a_{\mathcal{D}}^{3}}$.
Therefore the above equation now becomes

\begin{eqnarray}
\frac{\ddot{a}_{\mathcal{D}}}{a_{\mathcal{D}}} & = & \frac{g_{\mathcal{M}}^{3}t^{3\beta}}{a_{\mathcal{D}}^{3}}\frac{\beta(\beta-1)}{t^{2}}+\left(1-\frac{g_{\mathcal{M}}^{3}t^{3\beta}}{a_{\mathcal{D}}^{3}}\right)\frac{\alpha(\alpha-1)}{t^{2}}\nonumber \\
 &  & +2\frac{g_{\mathcal{M}}^{3}t^{3\beta}}{a_{\mathcal{D}}^{3}}\left(1-\frac{g_{\mathcal{M}}^{3}t^{3\beta}}{a_{\mathcal{D}}^{3}}\right)\left(\frac{\beta}{t}-\frac{\alpha}{t}\right)^{2}
\end{eqnarray}

From this equation we see that the global acceleration vanishes at
times given by 

\begin{eqnarray}
t^{3\beta}a_{\mathcal{D}}^{3} & = & \frac{1}{4\left(\beta-\alpha\right)g_{\mathcal{M}}^{3}}\left[\left(3\beta-\alpha-1\right)\right.\nonumber \\
 &  & \left.\pm\sqrt{\left(3\beta-\alpha-1\right)^{2}+8\alpha\left(\alpha-1\right)}\right]
\end{eqnarray}

This shows us that we get real time solutions for $\alpha\geq\frac{1}{3}\left[\left(\beta+1\right)+2\sqrt{2\beta\left(1-\beta\right)}\right]$.
If we now consider $\beta=0.5$ (its lowest possible value) then we
get $\alpha\geq0.971404521$ and if we consider $\beta=0.66$ (its
highest possible value) then we get $\alpha\geq0.999950246$. Hence
as stated earlier, for a particular value of $\beta$ we have a lower
limit on the value of $\alpha$.

In Fig. 1, for $\alpha_{j}<1$ the acceleration becomes negative in
the future for both cases of partitioning (curves (i) and (ii)). The
acceleration reaches a greater value and at a slightly later time
when $\mathcal{D}$ is partitioned into 100 overdense and underdense
subdomains (curve (ii)) and also becomes negative at an earlier time
as compared to the case where $\mathcal{D}$ is partitioned into 50
overdense and underdense subdomains (curve (i)). When $\alpha_{j}>1$
then we see that the acceleration curves for the two cases are almost
identical, with the maximum value being very slightly larger for partition
type (i) (curve (iii)). After reaching the maximum the acceleration
decreases and goes asymptotically to a small positive value. When
$\alpha_{j}<1$ then the first two terms of \eqref{eq:19} are negative,
but the last term, which is always positive, gains prominence as the
number of subdomains increases thus increasing the acceleration. When
$\alpha_{j}>1$ then only the first term in \eqref{eq:19} is negative
and hence the acceleration curves for the two partition cases (curves
(iii) and (iv)) are very similar, the only visible difference being
the slightly higher maximum value when $\mathcal{D}$ is partitioned
into a lower number of subdomains. 

In Fig. 2 we have illustrated the behaviour of the global acceleration
by taking narrow and broad ranges of values of $\beta_{j}$ and keeping
$\alpha_{j}<1$ and the same for all the curves. As seen in Fig. 1
here also the acceleration becomes negative in the future for all
the curves because we have $\alpha_{j}<1$ for all of them, but we
see that the difference between the acceleration curves for the two
partition cases is very small when we consider a narrow range of values
of $\beta_{j}$ (cuves (i) and (ii)) and the difference increases
considerably when we consider a broad range of values of $\beta_{j}$
(curve (iii) and (iv)). The acceleration attains a much greater value
when $\mathcal{D}$ is partitioned into a larger number of subdomains
and also becomes negative quicker. The reason for the latter behaviour
is that the broad range of values of $\beta_{j}$ makes the third
term in \eqref{eq:19} gain more prominence when we consider a larger
number of subdomains thus resulting in greater positive acceleration.

A similar comparison of the backreaction for the two models is presented
in Figs. 3 and 4 where we have plotted the backreaction density for
the duration over which the global acceleration is positive. We see
in these figures that the backreaction density is negative and from
the expression of $\Omega_{\mathcal{Q}}^{\mathcal{D}}$ and \eqref{eq:3}
it can be seen that the backreaction will be dominated by the variance
of the local expansion rate $\theta$. In Fig. 3 we see that for $\alpha_{j}<1$,
the backreaction density reaches a minimum, which is also greater
in magnitude, for partition type (ii) (curve (ii)) as compared to
partition type (i) (curve (i)). For $\alpha_{j}>1$ the curves for
the two cases are almost identical, the only difference being that
for partition type (i) (curve (iii)) the backreaction density reaches
a minimum of greater magnitude. In Fig. 4 we see that, just like the
acceleration curves, the difference between the backreaction plots
for the two partition cases is much smaller when we consider a narrow
range of values of $\beta_{j}$, but the difference beomes quite large
when we consider a broad range of values of $\beta_{j}$. The behaviour
of the backreaction as illustrated in Figs. 3 and 4 is quite similar
to the global acceleration, as seen in Figs. 1 and 2, and that is
expected because from \eqref{eq:14} we see that $\mathcal{Q}_{\mathcal{D}}$
is linearly proportional to the global acceleration.

\begin{figure}
\includegraphics[width=8.3cm]{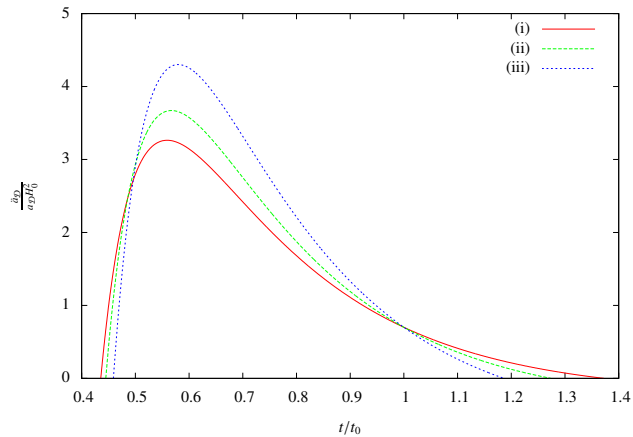}

\caption{We plot $\frac{\ddot{a}_{\mathcal{D}}}{a_{\mathcal{D}}H_{0}^{2}}$
vs. $t/t_{0}$ for various numbers of subdomains. In all the curves
we have $\alpha_{j}$ in the range 0.990 - 0.999, and $\beta_{j}$
in the range 0.58 - 0.60. For curve (i) we consider 100 overdense
and underdense subdomains, in (ii) 400 overdense and underdense subdomains
and in (iii) 500 overdense and underdense subdomains each.}

\end{figure}

In order to see how the global acceleration behaves based on the number
of subdomains we have in Fig. 5 plotted the global acceleration vs.
time for three partition cases where we consider (i) 100 overdense
and underdense subdomains, (ii) 400 overdense and underdense subdomains
and (iii) 500 overdense and underdense subdomains each. For all three
cases we have kept the range of values of $\alpha_{j}$ and $\beta_{j}$
the same and taken $\alpha_{j}<1$. It is clearly seen from the plot
that the global acceleration increases in magnitude as the number
of subdomains increases, and the maximum is obtained later in time
with increase in the number of subdomains. We also see that the acceleration
becomes negative faster when the number of subdomains increases.

\section{Conclusions}

To summarize, in this work we have performed a detailed analysis of
the various aspects of the future evolution of the presently accelerating
universe in the presence of matter inhomogeneities. The effect of
backreaction from inhomogeneities on the global evolution is calculated
within the context of the Buchert framework by considering the universe
to be divided into multiple underdense and overdense domains, each
evolving independently, in order to recreate the real universe more
accurately \cite{buchert1,buchert2,buchert3,weigand}. We analyze
the future evolution of the universe using the Buchert framework by
computing the global acceleration and strength of backreaction. We
show that the Buchert framework allows for the possibility of the
global acceleration vanishing at a finite future time, provided that
none of the subdomains accelerate individually (both $\alpha_{j}$
and $\beta_{j}$ are less than $1$).

Our analysis shows that if the $\beta_{j}$ parameters as distributed
over a narrow range of values and $\alpha_{j}<1$ then the global
acceleration reaches a greater maximum, when the number of subdomains
is larger, showing that the last term in \eqref{eq:19}, which is
always positive, has more prominence for a large number of subdomains.
This difference between the accelerations for the two partition cases
decreases even more when $\alpha_{j}>1$, because then only the first
term in \eqref{eq:19} has a negative contribution. However when we
consider a broad range of values of $\beta_{j}$ then the difference
between the accelerations for the two cases becomes much larger, the
acceleration being greater for a larger number of subdomains. The
cause for this is attributed to the dominance of the third term in
\eqref{eq:19} when we have a larger number of subdomains and a broad
range of values of $\beta_{j}$. We also saw that the behaviour of
the backreaction mimics the behaviour of the global acceleration,
and that is expected because as seen from \eqref{eq:14}, we have
$\mathcal{Q}_{\mathcal{D}}$ linearly proportional to the acceleration.

Our results indicate that backreaction can not only be responsible
for the current accelerated expansion, as shown in earlier works \cite{rasanen1,weigand},
but can also cause the acceleration to slow down and even lead to
a future decelerated era in some cases. In drawing this conclusion
it was not necessary for us to assume that the current acceleration
is caused by backreaction, and the acceleration could have been caused
by any other mechanism \cite{sahni}.

\end{document}